# Development of A Full-Scale Approach to Predict Overlay Reflective Crack

Zehui Zhu[a] and Imad L. Al-Qadi[a]

[a] *Illinois Center for Transportation, University of Illinois Urbana-Champaign, Rantoul, IL, USA*

Corresponding author: Zehui Zhu (zehuiz2@illinois.edu)

# Development of A Full-Scale Approach to Predict Overlay Reflective Crack

Resurfacing a moderately deteriorated Portland cement concrete (PCC) pavement with asphalt concrete (AC) layers is considered an efficient rehabilitation practice. However, reflective cracks may develop shortly after resurfacing because of discontinuities (e.g., joints and cracks) in existing PCC pavement. In this paper, a new accelerated full-scale testing approach was developed to study reflective crack growth in AC overlays. Two hydraulic actuators were used to simulate a moving dual-tire assembly with a loading rate of more than five-thousand-wheel passes per hour. A load cycle consists of three steps, simulating a tire approaching, moving across, and leaving a PCC discontinuity. Experiments were conducted to compare the reflective crack behavior of two overlay configurations. Both test sections were fully cracked in less than an hour. The initiation and propagation of reflective cracks were explicitly documented using crack detectors in conjunction with a camera. The proposed full-scale testing protocol offers a repeatable and efficient approach to systematically investigate the effects of various overlay configurations, thus enabling the identification of optimal design against reflective cracking.

**Introduction**

Pavement condition impacts mobility safety, vehicle-operating costs, and transportation infrastructure performance (Keenan et al., 2012). Approximately 61.8% of vehicle miles travelled on the US federal-aid highway system do not meet the established standard of good ride quality. Around 17.4% fail to qualify for acceptable ride quality (USDOT, 2021). Hence, rehabilitation is needed to restore the structural and functional capacity of deteriorated pavements. Common rehabilitation methods include reconstruction, resurfacing, and recycling. Resurfacing a moderately deteriorated Portland cement concrete (PCC) pavement with asphalt concrete (AC) layers is considered an efficient and common practice.

Reflective cracking is the most common distress observed in AC overlays. Because of discontinuities (e.g., joints and cracks) in existing PCC pavement, reflective cracks may develop shortly after resurfacing. Reflective cracks allow water to penetrate the pavement structure, leading to roughness and spalling (Son & Al-Qadi, 2014). The primary reflective cracking mechanisms are horizontal and differential movements caused by temperature/moisture changes and traffic loads (Huang, 2004). For example, pavement temperature changes periodically and varies at different depths. Periodic variations induce cyclic contraction and expansion, and contraction leads to relatively uniform tensile stress in the entire AC overlay. Because of accumulated horizontal movements at the joint, additional tensile stresses may add up in the AC overlay. In addition to high stresses at the top and bottom of an AC overlay, the temperature gradient causes warping of PCC slabs and aggravates the horizontal movements.

Traffic loads may induce both vertical and horizontal movements at the discontinuity. When a tire moves, a series of tensile and shear stresses occur at the bottom of the AC overlay. The underlying PCC conditions may significantly affect the magnitude of the resulting shear and tensile stresses. Hence, reflective cracking development may be governed by bending stresses, shear stresses, or both. The chance of developing reflective cracks increases at lower temperatures because AC is relatively more brittle and relaxes slower.

In the field of fracture mechanics, cracks are categorized into three modes (Anderson, 2017). Mode I, known as the opening mode, involves the application of principal loading normal to a crack plane, resulting in cracks that propagate perpendicular to the crack plane. Both thermal and traffic loading can induce mode I fracture. Mode II, characterized by cracks occurring in the in-plane shear direction, can be attributed to traffic loading, leading to mode II fracture due to the differential vertical movement of PCC slabs. Mode III entails cracks developing in the out-of-plane shear direction, potentially induced by lateral movement of concrete slabs, although such occurrences are rarely observed in AC overlays (Baek & Al-Qadi, 2008). Given that temperature and traffic loadings are typically applied concurrently in overlays, reflective cracking tends to develop in a mixed mode (Braham, 2008; Dave et al., 2010). While numerous studies have concentrated on mode I reflective cracking induced by thermal loading, comprehensive investigations into mixed-mode reflective cracking behaviours resulting from traffic loading have been comparatively limited (Xie & Wang, 2022).

Significant work has been conducted on reflective cracking mechanisms and potential methods to mitigate reflective cracks. In the field, reflective cracks may be studied once they reach the overlay surface. This method leads to inconclusive results because of high variability (Maurer & Malasheskie, 1989; Bennert & Maher, 2008; Bennert et al., 2009; Elseifi et al., 2011). Models using layer theory or finite-element (FE) analysis have been developed to simulate reflective cracking mechanisms (Jayawickrama et al., 1987; Elseifi & Al-Qadi, 2004; Minhoto et al., 2008; Baek & Al-Qadi, 2009; Dave & Buttlar, 2010; Lytton et al., 2010; Wang et al., 2018; Xie & Wang, 2022). Compared to closed-form linear elastic theory, FE analysis has the advantage of modeling complicated interlayer system geometry, moving traffic loads, and changing environmental conditions. Furthermore, relative to both field and laboratory studies, FE analysis offers a unique advantage by allowing for a detailed exploration of reflective cracking mechanisms. Specifically, it facilitates an in-depth examination of fracture modes within a given configuration through the computation of fracture parameters such as stress intensity factor for each mode. However, mechanistic models, in addition to computational cost, often lack extensive validation, calibration, and validation, limiting their abilities to guide AC overlay design in the field.

Full-scale testing allows researchers to simulate complex reflective cracking phenomena realistically under a controlled environment. However, only a few full-scale testing attempts have been made because of the relatively high cost and equipment constraints (Perez et al., 2007; Dave et al., 2010; Yin, 2015). Table 1 summarizes these efforts.

Table 1. Full-Scale Tests of Reflective Cracking in AC Overlay

| Study | Goal | Environment | Loading | Temperature | Duration | Instrumentation |
|---|---|---|---|---|---|---|
| Perez et al. | Verify & calibrate numerical model | Outdoor | Wheel load (65 kN) | 4~18℃ | ~30 days | • Strain<br>• Displacement<br>• Localized crack growth |
| Dave et al. | Investigate reflective crack relief interlayer | Outdoor | Wheel load (22~133 kN) | -10~2℃ | 8~10 days | • Displacement |
| Yin | Simulate thermal-induced reflective cracking | Indoor | Horizontal Pulling | -1.1~2.8℃ | ~45 days | • Strain<br>• Localized crack growth |

The key takeaways from the above-listed studies are summarized below:

- The measurement of reflective crack growth is a crucial task. However, limited success has been reported in the literature. A promising approach involves using a crack detector (CD), which is a single strand of copper wire glued onto the surface of the cross-section (Yin, 2015). However, it only provides localized measurements of crack growth due to the necessity of deciding the instrumentation locations before testing. The development of alternative techniques to monitor reflective crack growth is still an ongoing challenge.
- Full-scale testing is subject to high variability, as multiple variables could influence the outcome, including testing temperature, subgrade condition (e.g., moisture level), and AC density. For a meaningful comparison between overlay configurations, it is crucial to ensure that these variables are controlled.
- Testing duration ranged from 8 to 45 days. There is a need to develop an efficient protocol, allowing a thorough investigation of a wide variety of overlay configurations within a relatively short period.

In this paper, a new accelerated full-scale testing approach was developed to study the growth of reflective cracks. It uses two hydraulic actuators to simulate a moving dual-tire assembly with a loading rate of more than 5,000 wheel passes per hour. Experiments were conducted to investigate the reflective crack behavior of two overlay configurations. The initiation and propagation of reflective cracks were explicitly documented by deploying a robust instrumentation plan.

**Research Objective**

The main goal of this study was to develop an efficient full-scale testing protocol that allows a thorough investigation of reflective crack growth in AC overlays.

A large-scale testing device was designed and manufactured in-house. The loading configuration was carefully engineered to simulate a moving dual-tire assembly. Two test sections with different configurations were constructed and tested. Reflective crack growth was comprehensively measured using a robust instrumentation plan.

**Test Setup**

*Test Slab*

A 305-mm-thick fine sand subgrade layer was placed in a 3 × 1.8 m testbed and compacted by a vibratory compactor. The sand material is classified as poorly graded sand (SP) based on the Unified Soil Classification System (ASTM D2487, 2017). To evaluate the uniformity of density, lightweight deflectometer tests were performed at three locations after compaction according to ASTM E2835. An acceptable subgrade uniformity was achieved. Additionally, four 1.8 × 0.9 m neoprene rubber sheets were placed on the subgrade, with a total thickness of 95.25 mm. The incorporation of relatively less rigid neoprene rubber sheets emulated typical subgrade deterioration in the vicinity of the PCC joint, inducing stiffness degradation. Consequently, the PCC slabs exhibited increased deflection. The introduction of these neoprene rubber sheets resulted in a four-fold amplification of PCC vertical deflection, thereby elevating stress intensity and expediting the testing process.

Portland cement concrete slabs measuring 1.8 × 1.8 m square and 178 mm thick were cast following the procedures illustrated in Figure 1. Field-cured specimens underwent testing for compressive strength, flexural strength, and modulus of elasticity (AASHTO T 22, 2022; AASHTO T 177, 2017; ASTM C469, 2022). The 14-day compressive and flexural strengths were 25.7 MPa (3,728 psi) and 7.1MPa (1,024 psi), respectively, satisfying IDOT's minimum requirements of 24.1 MPa (3,500 psi) and 4.5 MPa (650 psi). The modulus of elasticity at 30 days was 30 GPa.

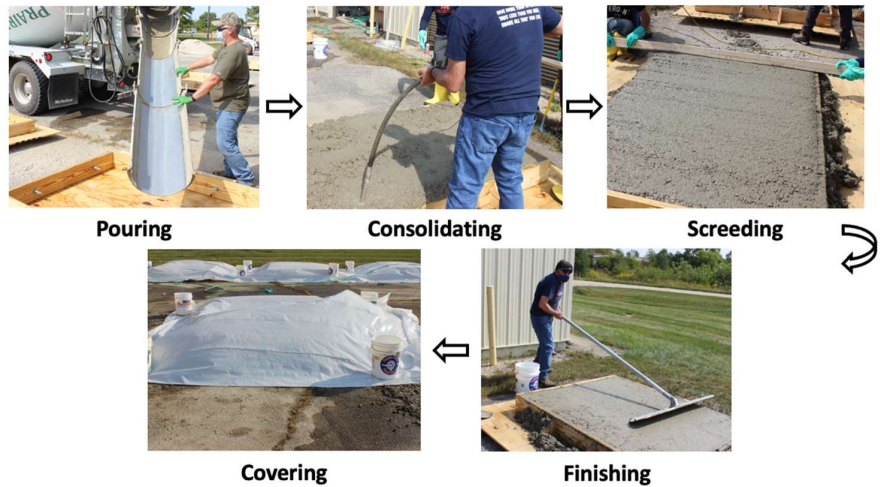

Figure 1. Portland cement concrete casting.

As presented in Figure 2, each test slab had four layers: a 305-mm fine sand subgrade, a 178-mm PCC slab, an AC binder course, and an AC wearing surface. Each test slab had a similar PCC slab and the same subgrade to ensure a fair comparison between AC overlay configurations.

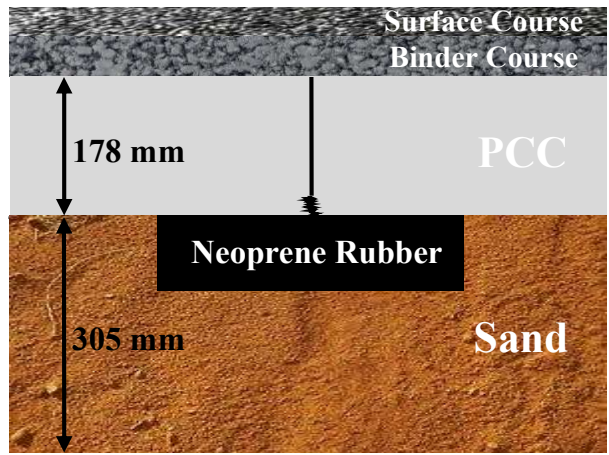

Figure 2. Schematic of test pavements.

As presented in Figure 3, constructing a test pavement involves the following steps:

- **PCC slab placement.** The PCC slabs were stored outside after casting. In preparation for testing, each slab was saw cut to around 152.4-mm depth, with a remaining unsawed portion of approximately 25.6-mm. The unsawed portion was left to provide sufficient strength for the concrete to withstand lifting during placement while maintaining the ability to easily crack after a few loading cycles. The joint was 8-mm-wide. The PCC slab was then brought indoors, and a crane was used to place the slab onto the test bed. The slab was finally situated on the subgrade. Note that the PCC slab center, 0.9 m wide, was in direct contact with the neoprene rubber sheet, while the rest was in direct contact with the sand subgrade.

- **Tack coat application.** An SS-1h tack coat was applied to the PCC surface two hrs before compaction. An application rate of 0.244 kg/m$^2$ was used in accordance with the Illinois Department of Transportation (IDOT) specifications. To accelerate debonding between PCC and binder course in the potential crack zone (i.e., center 0.6 m), the application rate was reduced by 50%. Research by Al-Qadi et al. (2008) has reported that a 50% decrease in application rate resulted in a minimum 70% reduction in interface shear strength.

- **Preheating.** Plant-mixed asphalt mixtures were stored at 10°C to prevent shelf aging (Al-Qadi et al., 2019). Subsequently, they underwent preheating to achieve a loose state in forced-draft ovens for approximately three hrs at 155°C, with the duration determined through experimental trials.

- **Remixing.** The AC materials were loaded into the mixing chamber of a recycler mixer using its conveyor. They were mixed and heated at 204°C for 30 min before discharging from the mixer. The selection of temperature and duration was based on experimental trials aimed at preventing prolonged exposure to excessively high temperatures.

- **Discharging.** The produced materials were discharged from the back of the recycler mixer and loaded into the steel bucket of a Bobcat skid-steer loader. The temperature of the discharged AC mix was carefully monitored throughout the production process. The mixing time was adjusted to ensure the discharge temperatures were close to 157°C for unmodified mixes and 163°C for polymer-modified AC mixes in accordance with IDOT specifications.
- **Laydown.** The AC mixture was dumped into the testbed, spread uniformly over the entire PCC slab, and leveled to approximately 1.25 times the target lift thickness (Habbouche, 2019).
- **Compaction.** A single-drum vibratory roller compactor was used to compact the lift. The vibration was turned off when compacting sand mix.
- **Surface course construction.** The following day, steps 2 to 7 were repeated to construct the surface course. The only difference was the tack coat application rate, where 0.122 kg/m$^2$ was used for the entire area following IDOT specifications.
- **Saw cutting.** A 300-mm of the AC overlay edge was cut and removed to create a smooth and clean cross-section to observe and measure reflective crack propagation.

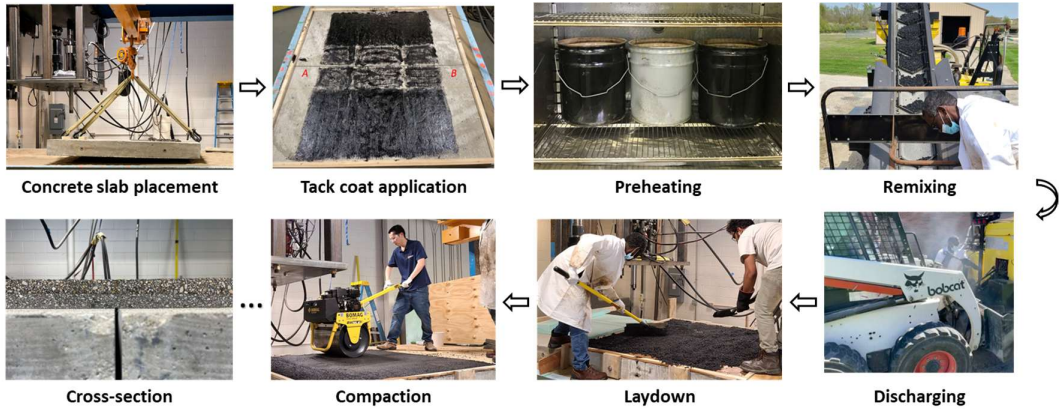

Figure 3. Test pavement construction.

*Instrumentation*

The main goal for slab instrumentation was to monitor reflective crack initiation and propagation. Additionally, measuring PCC slab movement was essential to evaluate the underlying condition and gain a better understanding reflective cracking mechanism. It should be noted that embedded instruments, such as strain gauges, were not utilized in this study to avoid interference with localized strain responses and initiated top-down cracking (Yin, 2015).

*Crack Detector*

The crack detector (CD) is a single strand of copper wire. Any erratic change in the recorded voltage indicates a discontinuity, meaning a crack propagates through (Yin, 2015). The CD has been effectively employed in both small-scale and full-scale tests to quantify crack growth (Wagnoner et al., 2005; Yin, 2015). Multiple copper wires were glued on the clean and smooth surface of the potential crack zone, as presented in Figure 4. A thin layer of coating was applied afterward for protection. The copper wires were connected to a laboratory power supply. A data acquisition system was used to record output signals throughout a test.

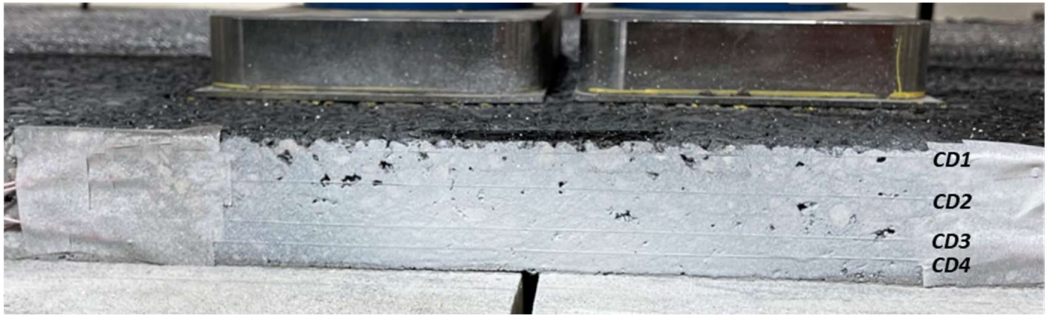

Figure 4. Crack detectors glued on overlay edge.

*Digital Camera*

Crack detectors only provide localized information and may not describe the complex cracking phenomenon. Additionally, the occurrence of hairline cracks or wire fatigue may give false indications. To overcome this limitation, an iPad was used to take pictures of potential crack zones every 100 loading cycles (i.e., 70 sec) during the testing, as presented in Figure 5.

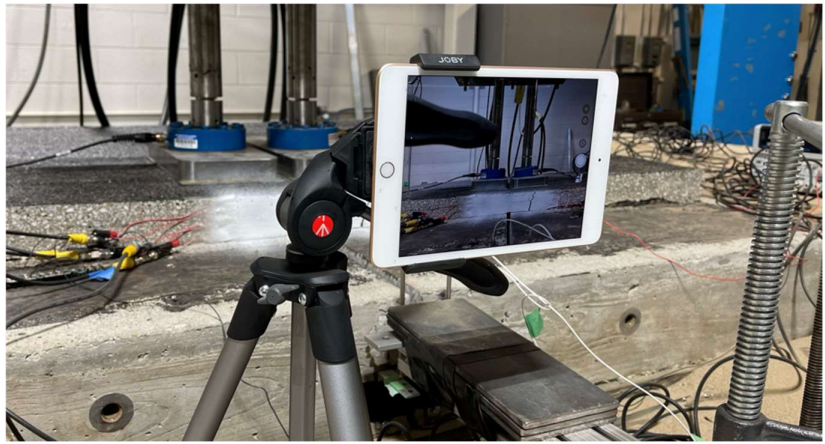

Figure 5. An iPad taking pictures during a test.

*Linear Variable Differential Transformer*

Two linear variable differential transformers (LVDTs) were used to measure the vertical movement of the PCC slab, as presented in Figure 6. LVDTs were placed at locations L and R to measure vertical deflections of the PCC slab on both sides of the joint.

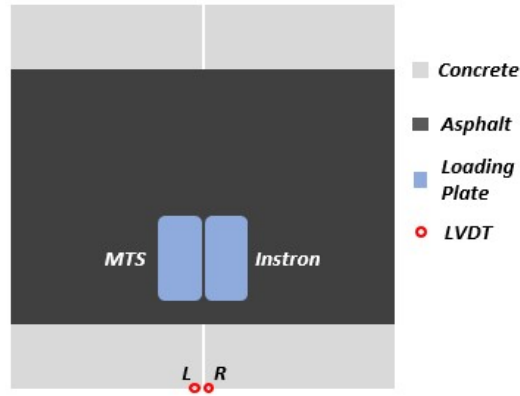

Figure 6. LVDT locations on the PCC slab.

*Testing*

*Loading System*

The loading system simulates a unidirectional dual-tire load. As presented in Figure 7, two rectangular loading plates measuring 381 × 178 mm were connected to two actuators (left and right) through rods and positioned on the AC overlay surface. Each loading plate represents a dual-tire assembly at 44.4 kN loading with a tire pressure of 650 kPa (Hernandez et al., 2017). The loading plate was positioned with a 152.4 mm spacing from the edge of the overlay. This distance was selected based on the experience reported by Perez et al. (2007) to facilitate the observation of crack propagation and prevent excessive creep of the material. The PCC joint was underneath the center of the gap between two loading plates, spaced at 25 mm. Each loading plate was associated with a load cell, which enabled the actuator to move the rod precisely to reach a desirable load. A pivot system was required to ensure a vertical load was applied on the AC. Both actuators were connected to the same hydraulic pump and received commands from the same controller system.

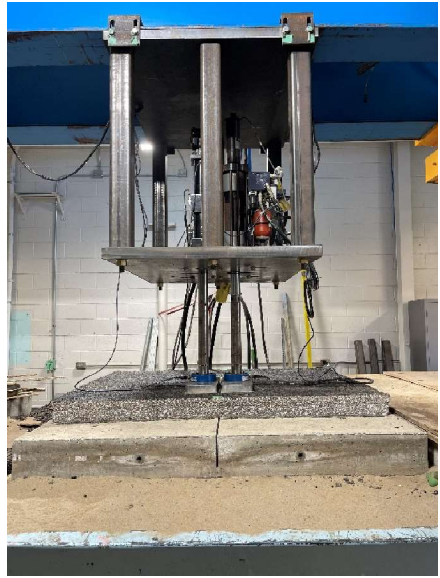

Figure 7. Hydraulic loading system.

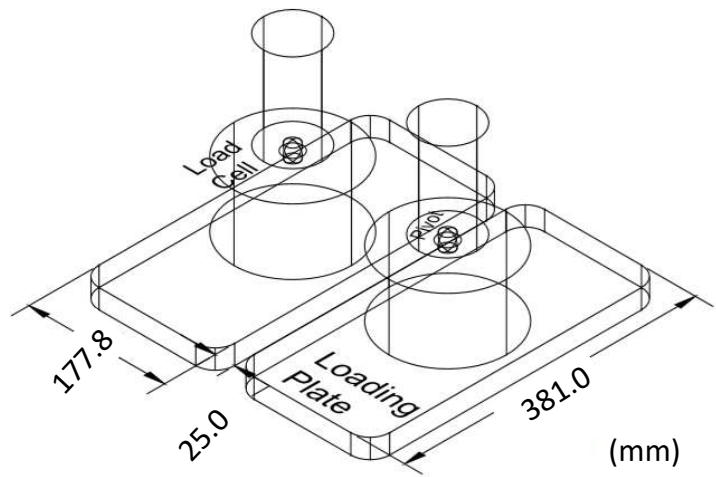

Figure 8. Loading plate configuration.

*Load Pattern*

Traffic-induced reflective cracking exhibits a mixed-mode behavior. As shown in Figure 9, when the tire approaches and leaves the PCC joint, there is in-plane shear (mode II). However, when the tire is on top of the crack, there is an opening (mode I), which is more dominant. Therefore, it is crucial to have a test setup that can replicate the actual tire pass. To achieve this, it is imperative to design a load pattern that considers both modes of fracture. As presented in Figure 10, a load cycle consists of three steps. In step 1 (tire approaching), only the left actuator applies loads; in step 2 (tire on top of the crack), both left and right actuators apply loads; and in step 3 (tire leaving), only the right actuator applies loads. Steps 1 and 3 mostly induce mode II (in-plane shear) fracture, while step 2 mainly leads to mode I (opening) fracture. A complete load cycle takes 0.42-sec, followed by a 0.28-sec rest until the next cycle starts. Figure 11 presents an example of two continuous load cycles.

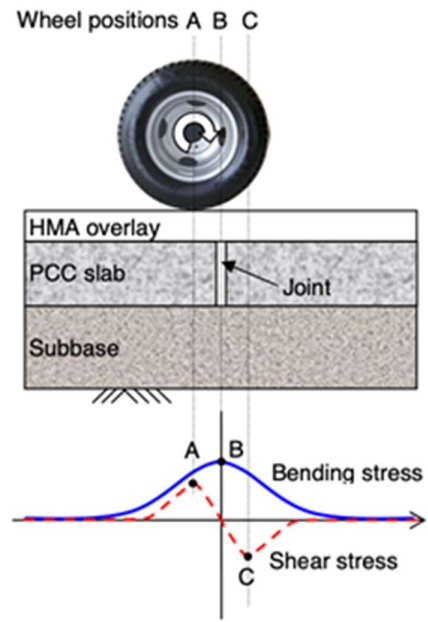

Figure 9. Reflective cracking mechanism caused by traffic loading.

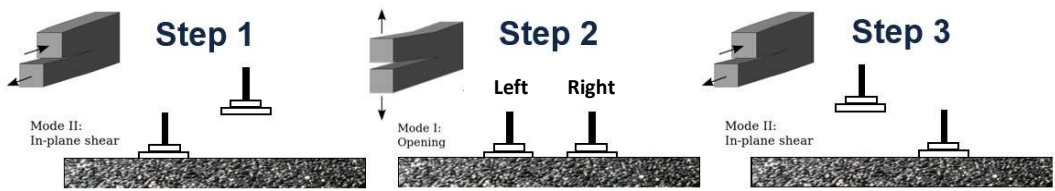

Figure 10. Steps of a load cycle.

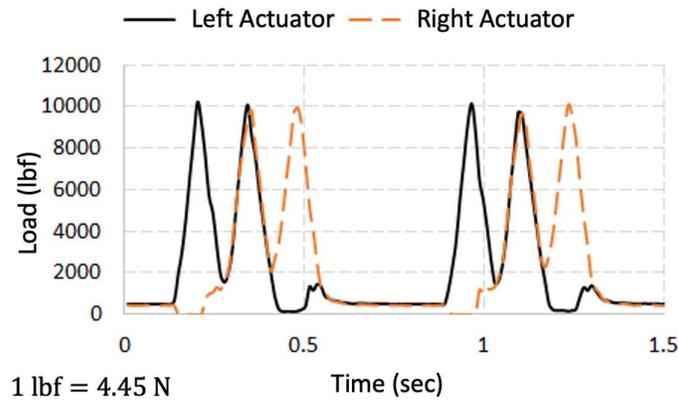

1 lbf = 4.45 N

Figure 11. Load pattern.

**Experimental Program**

Two test sections were constructed and tested. Table 2 summarizes their lift configurations, testing temperature, and duration. Section B differed from A by binder course only. Densities were measured from two cores taken close to the loading area after testing. All layers met IDOT's density requirements.

Table 2. Asphalt Concrete Lift Configurations and Testing Temperature.

| ID | Surface Course | | | Binder Course | | | Temperature (°C) | Duration (hr) |
|---|---|---|---|---|---|---|---|---|
| | Mix | Thickness (mm) | In-Place Density (%) | Mix | Thickness (mm) | In-Place Density (%) | | |
| A | IL-9.5 | 38.1 | 94.5 | IL-4.75 | 19.1 | 92.6 | 22.8 | 1.5 |
| B | | | 93.8 | IL-9.5FG | 31.8 | 92.3 | 23.5 | |

*Mix Design and Characterization*

Asphalt concrete mixtures were collected from various plants in Illinois. Table 3 summarizes the mix designs.

Table 3. Mix Design.

| Id | Mix Type | N-Design | NMAS* | AC (%) | Virgin Binder Grade | ABR+ (%) |
|---|---|---|---|---|---|---|
| IL-4.75 | Dense Graded | 50 | 4.75 mm | 8.2 | SBS PG 70-22 | 10.0 |

| IL-9.5 | Dense Graded | 70 | 9.5 mm | 6.1 | PG 58-28 | 29.3 |
| IL-9.5FG | Fine Graded | 90 | 9.5 mm | 5.9 | SBS PG 70-22 | 0.0 |

+ Asphalt binder replacement; * Nominal Maximum aggregate size

The Illinois-Flexibility Index Test (I-FIT) and Hamburg Wheel Tracking Test (HWTT) were performed to characterize each AC mix for cracking and rutting potential. Note that the materials were sampled and compacted using a gyratory compactor during overlay placement (AASHTO T 312, 2019). The reheating process in the forced-draft oven and recycler mixer could result in additional aging of the plant-mixed mixtures. This may increase brittleness and susceptibility to cracking, as well as enhanced stiffness and reduced susceptibility to rutting. In the field Material Transfer Vehicle (MTV) is used to reheat the material to placing temperature.

The I-FIT, which uses semi-circle beam geometry, is a simple and repeatable test that can differentiate AC mixes' cracking potential (Zhu et al., 2019). The test is carried out in load-line displacement control at a rate of 50 mm/min and a test temperature of 25 ℃. The fracture-mechanics-based index – Flexibility Index (FI) is the main output. A higher value of FI indicates a lower cracking potential (Ozer et al., 2016; AASHTO T 393, 2022). The HWTT evaluates AC's susceptibility to permanent deformation. Each of the two sets used in the machine compromises two cylindrical specimens with a 150 mm diameter and a 62 mm height. Each specimen was placed in a mounting tray, submerged in a warm water bath at 50°C, and subjected to repetitive steel-wheel sinusoidal loading. The steel wheel weighs 71.67 kg and is applied at 52 passes/min across each set. The main output is the rut depth formed on the specimens after a certain number of passes. Smaller rut depth is an indicator that the mix is more rutting resistant (AASHTO T 324, 2019).

Figure 12 shows the balance mix design (BMD) interaction plot between rut depth and FI. IL-4.75 demonstrated good cracking and rutting resistance by falling exclusively within the most desirable quadrant, visually distinguished by its green shading. Conversely, IL-9.5 exhibited high cracking and rutting potential.

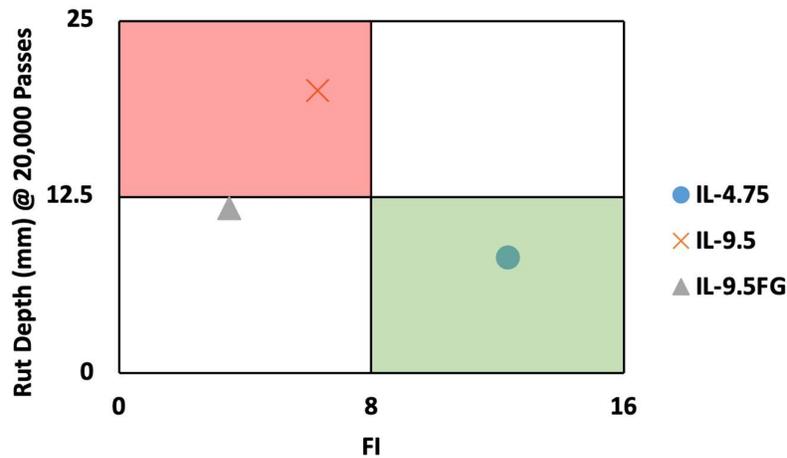

Figure 12. 2-D interaction plot between rut depth and FI.

*Full-Scale Test Results*

*Concrete Slab Movement*

The LVDTs were placed on both sides of the joint to measure the movement of the PCC slab. Figure 13 compares the PCC vertical deflections of sections A and B of a loading cycle at the start of the test. Under equivalent loading conditions, it was observed that section B's PCC displayed a slight increase in deflection compared to section A, with similar load transfer efficiencies noted – 43.3% for section A and 52.0% for section B. The similar PCC deflections between sections A and B indicate that the underlying conditions of the overlays were similar. Hence, differences in reflective cracking behavior could be attributed primarily to AC overlay configurations. This finding confirmed the repeatability of the test protocol, as variables such as testing temperature, subgrade condition, and AC density, which are typically difficult to control in full-scale testing, were well-controlled in the proposed protocol.

The left half of the PCC slab had less deflection under loading step 2, which was most likely due to non-flat PCC bottoms resulting from wood frame distortion during PCC casting.

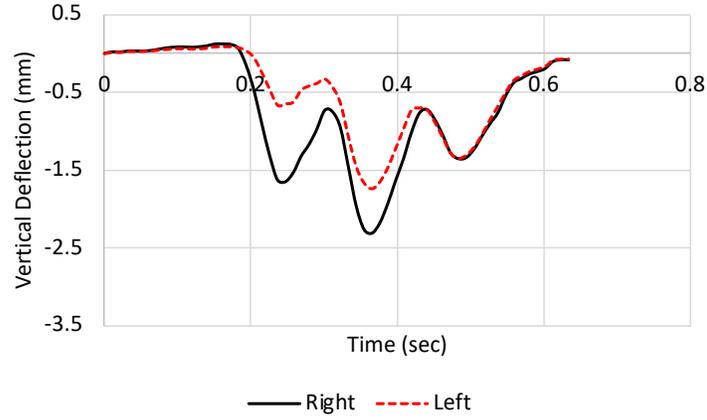

*(a) Test section A*

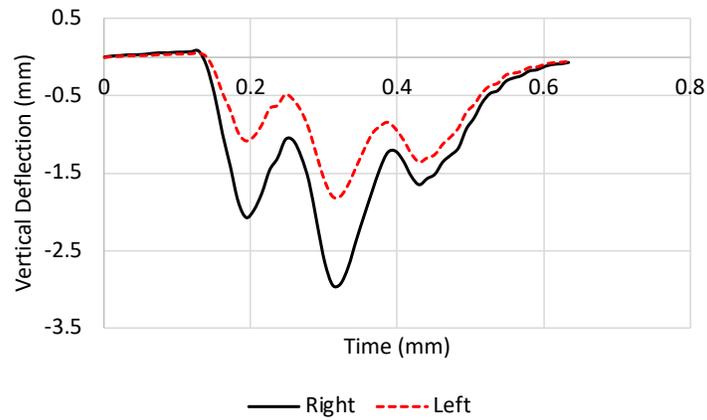

*(b) Test section B*

Figure 13. Vertical deflections of the PCC slab.

*Reflective Crack Growth*

Crack detectors (i.e., a single strand of copper wire) and a camera were used to monitor reflective crack growth. Implementing crack detectors in conjunction with a camera enabled a mutual validation process, ensuring precise crack growth measurements.

Figure 14 presents images of the overlay cross-section at various loading cycles. Reflective crack growth was documented. A reflective crack initiated at the bottom of the overlay for both sections and propagated upwards. The crack initiated on the right of the joint, near the end of the debonding area. The crack path did not follow a pure vertical trajectory, indicating a mixed-mode fracture. Although the binder course of section B was 12.7 mm thicker than section A, its relatively low binder content and low FI underscored the performance. Despite a relatively thicker structure, section B failed faster than section A.

Figure 15 shows the reflective crack length versus the number of loading cycles. The average crack growth rate for sections A and B was 11.9 mm and 18.8 mm per one thousand cycles, respectively. The difference primarily is due to crack initiation. As a stress-absorbing layer, the high-FI IL-4.75 sand mix significantly delayed the reflective crack initiation and efficiently enhanced section A to control reflective cracking. These results agreed with those reported in the literature (Baek et al., 2011; Greene et al., 2016). Because sections A and B shared the same surface course, the crack growth rate was similar within the surface layer.

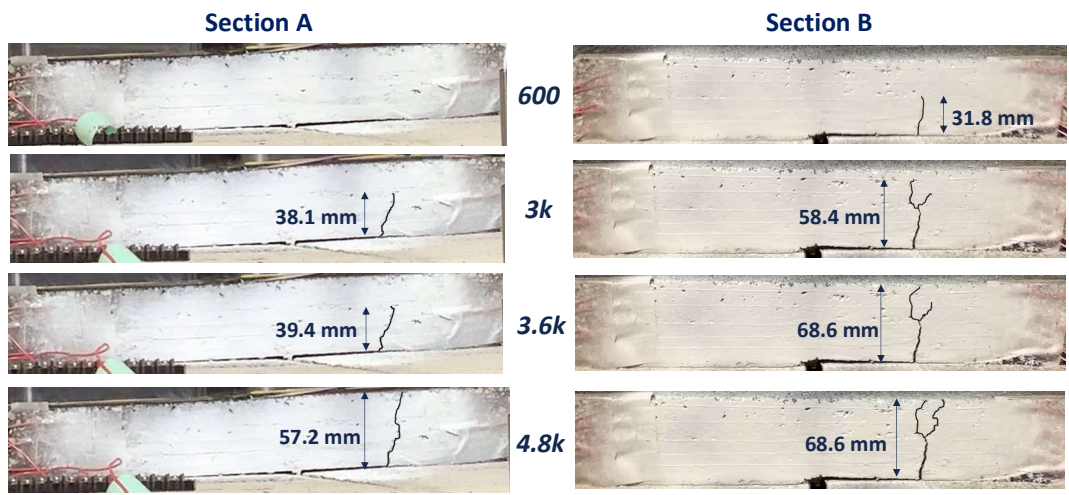

*For better visualization purpose, cracks were marked with black lines.*

Figure 14. Reflective crack growth.

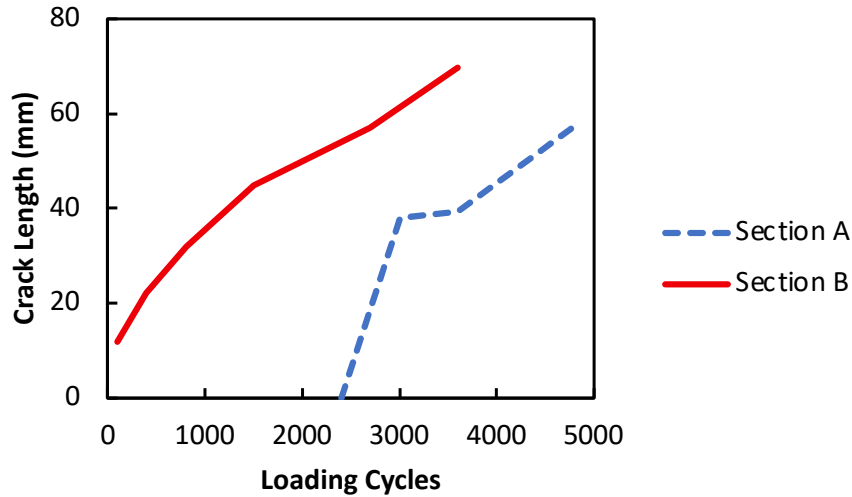

Figure 15. Reflective crack length versus loading cycles.

**Summary and Conclusions**

Resurfacing a moderately deteriorated PCC pavement with AC layers is an efficient rehabilitation practice. However, reflective cracks may develop shortly after resurfacing due to discontinuities (e.g., joints and cracks) in existing PCC pavement. In this paper, a new accelerated full-scale testing protocol was developed to quantify reflective crack growth in AC overlays. It uses two hydraulic actuators to simulate a moving dual-tire assembly with a loading rate of more than 5,000 wheel passes per hour. A load cycle consists of three steps. In the first step, the left loading plate applies a load while the right plate applies almost zero or no loads. Step 1 represents the scenario of a tire approaching the crack. In the second step, both plates apply the load simulating the tire on top of the crack. The third step is the reverse of the first step: only the right plate applies a load, and the left plate does not. Step 3 models the tire leaving the crack. Experiments were conducted to compare the reflective crack behavior of two overlay configurations. Both test sections were fully cracked in less than an hour. The initiation and propagation of reflective cracks were explicitly documented by deploying a robust
21

instrumentation plan.

This study concluded the following:

- A repeatable full-scale reflective cracking testing protocol is introduced. The experiment setup could effectively control variables such as testing temperature, subgrade condition, and AC density, which are often challenging to regulate in full-scale testing. The comprehensive control of these factors allowed for the systematic investigation of various overlay configurations.
- The proposed test protocol is efficient; test sections were fully cracked in less than one hour (5,000 loading cycles). Hence, a rapid identification of optimal overlay configurations against reflective cracking could be obtained.
- The implementation of crack detectors in conjunction with a camera-enabled a mutual validation process, which ensured precise crack growth measurements.
- As a stress-absorbing layer, the high-FI IL-4.75 sand mix significantly delayed the reflective crack initiation and efficiently enhanced the overlay's resistance against reflective cracking. Although the binder course of section B was 12.7 mm thicker than section A, its relatively low binder content and low FI underscored its performance. Hence, overlay material characteristics and layer thickness are both important to control reflective cracking.

**Limitations and Recommendations**

The following are the limitations of this study and relative suggestions for future research:

- Asphalt concrete exhibits varying modulus values at different load frequencies due to its viscoelastic nature. Small-scale laboratory investigations and numerical modelling have demonstrated the substantial influence of load frequency on the reflective cracking behaviour of AC overlays (Gonzalez-Torre et al., 2015). While this study focused on a specific load frequency, it is recommended to explore the impact of load frequency on reflective crack growth using the proposed full-scale methodology.

- The proposed approach primarily simulates traffic-induced mixed-mode reflective cracking. It is imperative to recognize that under varying environmental and overlay conditions, alternative mechanisms, notably thermal loading, may exert dominant effect. An efficient and cost-effective full-scale approach capable of simulating the combined effects of both traffic load and thermal load should be investigated.

- A full-scale approach could demonstrate the influence of distinct fracture modes on reflective crack growth in mixed-mode scenarios. However, finite element analysis is better suited for providing quantitative results. The results obtained in this study could be used to validate (and possibly calibrate) numerical simulations.

**Acknowledgment**

This publication is based on the results of ICT-R27-204, Optimized Hot-mix Asphalt (HMA) Lift Configuration for Performance. ICT-R27-204 was conducted in cooperation with the Illinois Center for Transportation; the Illinois Department of Transportation; and the U.S. Department of Transportation, Federal Highway Administration. Special thanks to the ICT research staff, Greg Renshaw, Uthman Mohamed Ali, Marc Killion, and Mohsen Motlagh for their input and support during this study. The support of the following colleagues is truly appreciated: Aravind Ramakrishnan, Ghassan Chehab, Izak Said, Punit Singhvi, Lama Abufares, and Akash Bajaj. The contributions of the technical review committee are acknowledged;

especially the input of the co-chairs, John Senger and Loura Heckle. The contents of this paper reflect the view of the authors, who are responsible for the facts and the accuracy of the data presented herein. The contents do not necessarily reflect the official views or policies of the Illinois Center for Transportation or the Illinois Department of Transportation. This paper does not constitute a standard, specification, or regulation.24